\documentclass[prd,aps,floats,12pt,nofootinbib]{revtex4}

\usepackage{amsmath}
\usepackage{amssymb}

\usepackage{epsfig}

\newcommand{\beq}{\begin{equation}}
\newcommand{\eeq}{\end{equation}}

\newcommand{\ber}{\begin{eqnarray}}
\newcommand{\eer}{\end{eqnarray}}

\def\beq{\begin{equation}}
\def\eeq{\end{equation}}

\def\ber{\begin{eqnarray}}
\def\eer{\end{eqnarray}}

\begin{document}

\title{Is it time to go beyond $\Lambda$CDM universe?}

\author{Anto  I. Lonappan}
\email{antoidicherian@gmail.com} 
\affiliation{Department of Physics, S.B College, Changanassery, Kottayam, Kerala, 686101, India} %

\author{Ruchika}
\email{ruchika@ctp-jamia.res.in} 
\affiliation{Centre for Theoretical Physics, Jamia Millia Islamia, New Delhi-110025, India} %

\author{Anjan A Sen}
\email{aasen@jmi.ac.in} 
\affiliation{Centre for Theoretical Physics, Jamia Millia Islamia, New Delhi-110025, India}

\begin{abstract}
Concordance $\Lambda$CDM universe is the simplest model that is consistent with a large variety of cosmological observations till date. But few recent observations indicate inconsistencies in $\Lambda$CDM model. In this paper, we consider the combination of  recent SnIa+Bao+Cmb+Growth+$H(z)$+$H_{0}$ measurements to revisit the constraints on the dark energy evolution using the widely studied CPL parametrisation for the dark energy equation of state. Although the reconstructed behaviour for the dark energy equation of state confirms the inconsistency of $\Lambda$CDM at $95\%$ confidence level, the reconstructed $Om$ diagnostic which is a {\it null test} for $\Lambda$CDM,  still allows the concordance $\Lambda$CDM behaviour with a lower range of $\Omega_{m0}$ than that obtained by Planck-2015. {\it This confirms that $\Lambda$CDM is still the best choice for the dark energy model}. We also measure the parameter $S = \sigma_{8}\sqrt{\Omega_{m0}/0.3} = 0.728 \pm 0.023$ which is consistent with its recent measurement by KiDS survey. The confidence contour in the $\Omega_{m0}-\sigma_{8}$ parameter plane is also fully consistent with KiDS survey measurement.
\end{abstract}
\maketitle

For nearly two decades, the most pressing problem in cosmology is to explain the late time acceleration of the universe. Almost all the present and future cosmology missions are dedicated to address this issue. After the Planck-2015 data \cite{ade} for anisotropy in the cosmic microwave background radiation (CMB) together with other cosmological data from Supernova type-Ia (SNIa) \cite{jla}, Baryon Acoustic Oscillations (BAO) observation \cite{bao} in the large scale structures in the universe as well as the HST measurement of the Hubble parameter \cite{hst}, the concordance $\Lambda$CDM universe is shown to be consistent with this combined datasets. Being the simplest model in explaining the late time acceleration, this makes $\Lambda$CDM universe the clear winner among various dark energy models, although the theoretical issues such as the fine tuning and cosmic coincidence problems are still far from being solved and will keep the theoreticians busy in near future ( see \cite{de} for nice review on dark energy).

Despite the success of $\Lambda$CDM universe to explain a large variety of cosmological observations, there are recent evidences to contradict this success \cite{discrep}. Recently the latest model independent measurement of $H_{0}$ (hereafter Riess16) \cite{riess},  has more than $3\sigma$ deviation from the Planck-2015 measurement of the same for a $\Lambda$CDM universe. More recently, the KiDS survey \cite{kids} has found a discrepancy in growth measurement at the level of $2.5\sigma$ compared to the measurement by Planck-2015 for $\Lambda$CDM model. Just recently, Valentini et al \cite{valen} have shown that the $\Lambda$CDM model is inconsistent at $95\%$ confidence level with the Planck-2015 + Riess16 dataset. This clearly motivates people to revisit the constraint on dark energy behaviour.

Going beyond the $\Lambda$CDM universe where the dark energy density is constant throughout the evolution of the universe, there are several approaches to model the dark energy evolution. One can assume simply a constant negative equation of state for the dark energy ( like those arise from a network of strings or domain walls \cite{topo}). The other choice is to consider a minimally coupled canonical scalar field slowly rolling over a sufficiently flat potential to mimic a negative equation of state. This is similar to the inflaton field in the early universe with the difference that in this case the scalar field evolves at a much lower energy scales. In literature these are called {\it quintessence} field  \cite{quint}. Subsequently Caldwell and Linder \cite{lindcald} showed that such  quintessence fields can be further divided into categories called {\it freezing} and {\it thawing} quintessence. The freezing model has an initial fast roll phase where they mimic the background radiation or matter behaviour; later on, the equation of state decreases and asymptotically approaches $w=-1$ to initiate the late time acceleration. For the thawing models, the field is initially frozen at the flat part of the potential due to large Hubble friction and behaves like $w=-1$ cosmological constant; subsequently the Hubble friction decreases and the field starts rolling and the equation of state starts increasing with time. Generalization of these models to non-canonical \cite{noncan}, non-minimal \cite{nonmin} as well as phantom ($w < -1$) \cite{phantom} cases have also been studied extensively. In a recent work, Dhawan et al. \cite{dhawan} have studied the bayesian evidences for a variety of dark energy models using CMB/BAO+SnIa data.

On the other hand, given the large number models in the literature, it is rather difficult to confront all individual models to the observational data. The more economical way is to construct parametrization of the dark energy equation of state $w$ as a function of redshift or scale factor containing a minimal set of parameters that describes a wide set of dark energy models and then confront such parametrization to the observational data. One such widely used parametrization is the Chevallier-Polarski-Linder (CPL) parametrization \cite{cpl} which was been widely used by  all the recent cosmological observations including Planck-2015 to put constrain on the cosmological parameters.

In this paper, we use this CPL parametrization to constrain the dark energy behaviour using the current observations from CMBR, SNIa, BAO, the growth of the matter fluctuations and the measurement of the Hubble parameter at different redshifts including Riess16 data.

To start with, we first describe the CPL parametrization for the dark energy equation of state which has the following form:

\begin{equation}
w(a) = w_{0} + w_{a} (1-a) = w_{0} + w_{a} \frac{z}{1+z},
\end{equation} 

\noindent
where $w_{0}$ and $w_{a}$ are the two parameters of the model describing the equation of state at present ($a=1$) and the variation of the equation of state at present respectively. From the infinite past $(a=0)$ till the present $(a=1)$, the equation of state varies between $w_{0} + w_{a}$ and $w_{0}$. Using this form for the equation of state into the energy conservation equation for the dark energy, one can easily get the variation of the dark energy density as

\begin{equation}
\rho_{de} \propto a^{-3(1+w_{0}+w_{a})} e^{-3w_{a}(1-a)}.
\end{equation}

\noindent
Interestingly this simple form of the equation of state fits a wide range of scalar field dark energy behaviours including the supergravity motivated SUGRA model for dark energy \cite{sugra}. It is easy to check that for $w_{0} > -1$ and $w_{a}>0$, the dark energy remains non-phantom $(w(a) > -1)$ throughout the cosmological evolution. Otherwise it behaves like phantom $(w(a) < -1)$ at some point in time in the cosmological evolution. As discussed earlier, one of the most studied example of dark energy is quintessence model with a time dependent scalar field. Depending on the initial conditions and the nature of the potential for the scalar field, quintessence field can either have freezing behaviour or thawing behaviour. In an interesting paper, Caldwell and Linder  \cite{lindcald} have obtained two restricted regions in the $w_{0}-w_{a}$ parameter space, where the quintessence field behaves either as freezing model or as thawing model.

With this, the $0-0$ component of the Einstein equation for a spatially flat FRW universe that contains matter and dark energy is given by

\begin{equation}
H^{2}(a) = H_{0}^2 \left(\Omega_{m0} a^{-3} + \Omega_{r0} a^{-4} + (1-\Omega_{m0}-\Omega_{r0}) a^{-3(1+w_{0}+w_{a})} e^{-3w_{a}(1-a)} \right).
\end{equation} 

\noindent
Here $\Omega_{m0}$ and $\Omega_{r0}$ are the present day density parameter for matter (that includes baryons and cold dark matter) and radiation respectively and $H_{0}$ ($100h \hspace{1mm} Km/sec/Mpc$) is the present day Hubble parameter. This expression for the Hubble parameter $H(a)$ with five parameters $h, w_{0}, w_{a}, \Omega_{m0}$ and $\Omega_{r0}$ is sufficient to calculate all the observable quantities related to the background cosmology.

To distinguish different dark energy models among themselves and also with the $\Lambda$CDM model, there is a very useful diagnostic, called {\it Om diagnostic} proposed by Sahni et al. \cite{om}. It is defined as

\begin{equation}
Om(z) = \frac{(\frac{H(z)}{H_{0}})^{2}-1}{(1+z)^3-1}.
\end{equation}

\noindent
It is not difficult to see that for $\Lambda$CDM model, $Om(z)$ is constant throughout the evolution of the universe. It provides powerful {\it null test} for $\Lambda$CDM model whereby a evolving $Om(z)$ confirms a non-$\Lambda$CDM model.

To study the growth of matter fluctuations on sub-horizon scales where dark energy behaves as a smooth component, we take the linearised equation for growth of matter density contrast under Newtonian approximation as:

\begin{equation}
\ddot{\delta_{m}} + 2 H \dot{\delta_{m}} - 4\pi G \bar{\rho_{m}} \delta_{m} = 0,
\end{equation}

\noindent
where $\delta_{m}$ is the matter density contrast, $\bar{\rho_{m}}$ is the background matter energy density and ``{\it overdot}'' represents the derivative with respect to the cosmological time $t$. $H$ is the Hubble parameter given by equation (3). We also define linear growth rate as:

\begin{equation}
f = \frac{d\log\delta_{m}}{d\log a}.
\end{equation}

The quantity $f(z)\sigma_{8}(z)$,  where $\sigma_{8}(z)$ is the rms fluctuation of linear density field $\delta_{m}$ within a box size of $8 h^{-1}$ Mpc, is a model independent estimator of the observed growth history in the universe. On sub-horizon scales where dark energy behaves like a smooth component, one can write \cite{luca} 

\begin{equation}
f(z) \sigma_{8}(z) = \sigma_{8} \frac{\delta_{m}^{'}}{\delta_{m}(z=0)},
\end{equation}

\noindent
where ``{\it prime}'' denotes differentiation with respect of $\log (a)$ and $\sigma_{8} = \sigma_{8} (z=0)$.

\vspace{3mm}
\begin{table}
\begin{tabular}{|c|c|c|c|c|c|}
\hline $\Omega_{m0}$ & $w_{0}$ & $w_{a}$ & $h$ & $\Omega_{r0}$ & $\sigma_{8}$ \\ 
\hline $0.283\pm 0.009$ & $-0.91\pm 0.11$ & $-0.70^{+0.5}_{-0.42}$ & $0.702\pm 0.0074$ & $(4.68\pm 0.20) \times 10^{-5}$ & $0.749\pm 0.023$ \\ 
\hline 
\end{tabular} 
\caption{The 1D marginalised $68\%$ confidence intervals for the model parameters.}
\end{table}

With this, we use the latest observational data to constrain the parameters in our model and subsequently constrain the behaviour of the dark energy. We use the following observational data for our analysis:

\begin{itemize}

\item 

We use the measurements of the luminosity distance of SNIa from the ``{\it Joint Light Curve Analysis (JLA)}''  taken from SDSS and SNLS catalogue \cite{jla}.

\item 

We use the combined BAO/CMB constraints on the angular scales of the BAO oscillations in the matter power spectra as compiled by Giostri et al \cite{giostri}.

\item 

We use the measurement of $f(z)\sigma_{8}(z)$ by various galaxy surveys as compiled by Basikalos et al \cite{growth}.

\item

We use the acoustic scale and CMB shift parameters as measured by Planck-2015 observations \cite{ade1}.

\item

We use the measurement of Hubble parameter as a function of redshift as compiled by Farooq et al \cite{hubble}.

\item

We also use the latest measurement of $H_{0}$ by Reiss et al \cite{riess} (Riess16).

\end{itemize}

With these set of cosmological data, we use publicly available Markov Chain Monte Carlo (MCMC) package ``{\it emcee}'' \cite{emcee} to put constrain on the parameters in our model. To obtain the covariance matrices from the MCMC sample chains, we use the publicly available python package ``{\it GetDist}'' \cite{getdist}.

\begin{figure*}[!htb]
\begin{center}
\resizebox{260pt}{220pt}{\includegraphics{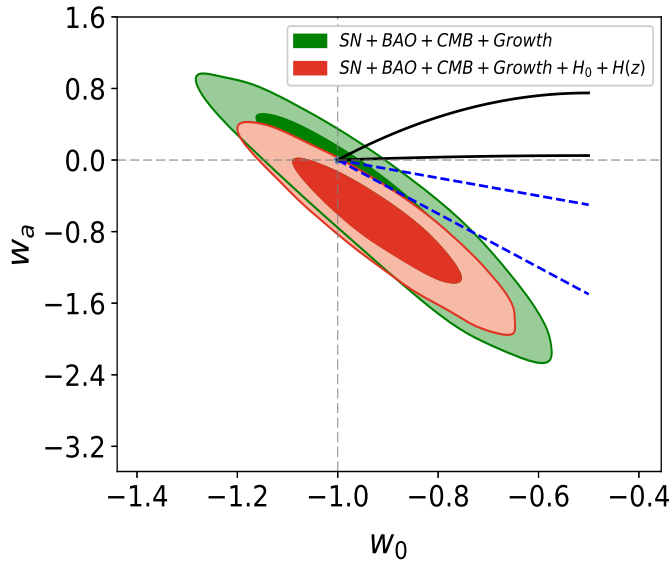}} 
\end{center}
\caption{$68\%$ and $95\%$ confidence contours in the $w_{0}-w_{a}$ parameter plane using two different set of observational data. The region inside the dashed line, the quintessence field behaves like thawing model whereas the region inside the solid lines, the quintessence field behaves as freezing model (see \cite{lindcald}).} 
\end{figure*}

\begin{figure*}[!htb]
\begin{center} 
\resizebox{260pt}{220pt}{\includegraphics{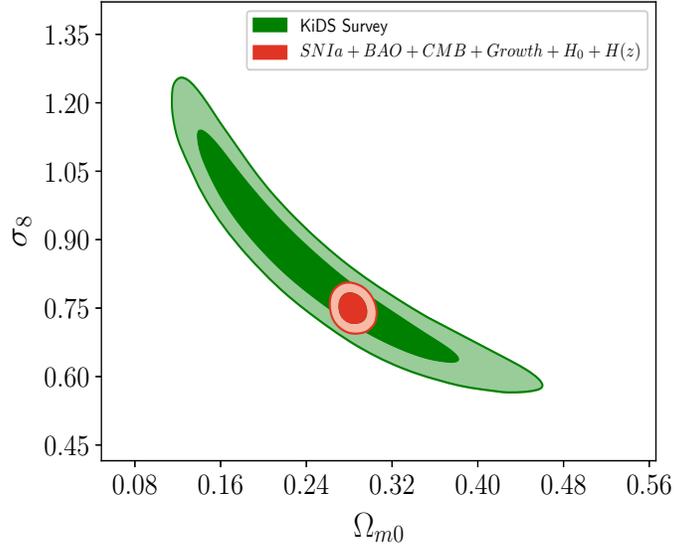}} 
\end{center}
\caption{$68\%$ and $95\%$ confidence contours in the $\Omega_{m0}-\sigma_{8}$ parameter plane. The KiDS survey plot is based on data products from observations made with ESO Telescopes at the La Silla Paranal Observatory under programme IDs 177.A-3016, 177.A-3017 and 177.A-3018 ( see text for detail references).} 
\end{figure*}

In table 1, we quote the 1-D marginalised $68\%$ confidence interval for our model parameters. As one can see, the constrain on $\sigma_{8}$ is significantly lower than the Planck-2015 measured value of $\sigma_{8}$ for $\Lambda$CDM model. This is due to the inclusion of growth data which are already in tension with Planck-2015 measurement.  This measurement is in tension with the $\sigma_{8}$ measurement by CFHTLenS tomographic weak lensing survey at $1.8\sigma$ which is not significant. We also measure the parameter $S = \sigma_{8} \sqrt{\Omega/0.3}$ which is $0.728 \pm 0.023$. This is consistent with that measured by Kilo Degree Survey (KiDS) \cite{kids}.  Our measured $h$ is also consistent with independently measured value for $h$ by HoLiCOW project \cite{holicow} using three strong lenses system.

In figure 1, we show the confidence contours in the $w_{0}-w_{a}$ parameter space. In the same contour we also show the regions obtained by Caldwell and Linder, where the scalar field quintessence models behaves either as freezing or as thawing model. As one can see, without the data for the Hubble parameter, the cosmological constant ($w_{0} = -1 \hspace{1mm} \& \hspace{1mm} w_{a} = 0$) is perfectly consistent. But once we add the $H(z)$ and $H_{0}$ data, the cosmological constant model sits at the edge of the $95\%$ confidence region. Moreover the freezing models are now ruled out at $95\%$ confidence level whereas the thawing models are just marginally allowed at $95\%$ level. Given that the minimally coupled, canonical scalar quintessence field can have either thawing or freezing behaviour, the figure 1 shows that these kind of scalar field models are practically ruled out at $95\%$ confidence level except a very tiny region for thawing models that is allowed at $95\%$ confidence interval.

In figure 2, we show the confidence contour in the $\Omega_{m0}-\sigma_{8}$ plane for the SNIa+BAO+CMB+Growth+H(z)+$H_{0}$ data.  In this plot we also show the same confidence contour as obtained by KiDS tomographic weak lensing survey. For this, we use cosmic shear measurements from the Kilo-Degree Survey (Kuijken et al. 2015 \cite{kuijken}, Hildebrandt et al.  2017 \cite{kids}, Fenech Conti et al. 2016 \cite{conti})( KiDS). The KiDS data are processed by THELI (Erben et al. 2013  \cite{erben}) and Astro-WISE (Begeman et al.  2013 \cite{begeman}, de Jong et al.  2015 \cite{dejong}). Shears are measured using lensfit (Miller et al. 2013 \cite{miller}), and photometric redshifts are obtained from PSF-matched photometry and calibrated using external overlapping spectroscopic surveys (see Hildebrandt et al. 2016 \cite{hild}). As one can see, our measurement is fully in agreement with the KiDS result.

\begin{figure*}[!htb]
\begin{center} 
\resizebox{260pt}{220pt}{\includegraphics{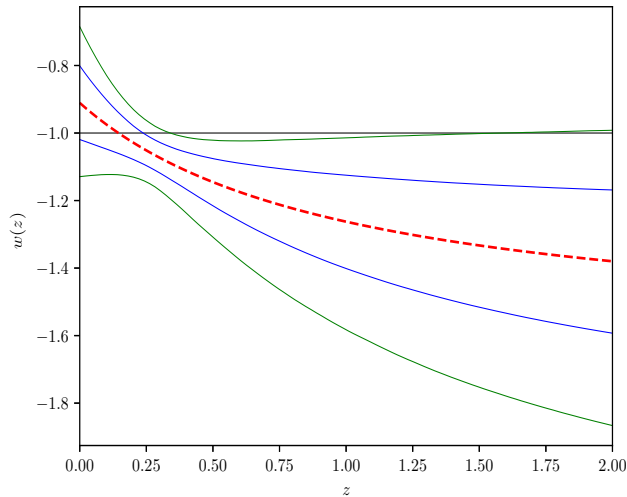}} 
\end{center}
\caption{$68\%$ and $95\%$   confidence region for the reconstructed $w(z)$ with SNIa+BAO+CMB+Growth+H(z)+$H_{0}$ data. The dashed line represent the best fit behaviour. The horizontal solid line represent the $w=-1$ cosmological constant behaviour.} 
\end{figure*}

We should mention that the two parameters $w_{0}$ and $w_{a}$ are related to the dark energy equation of state at present ($z=0$). Hence the confidence contour shown in figure 1, strictly represents dark energy property at present. To constrain the evolution of the dark energy equation of state $w(z)$, we need to reconstruct $w(z)$ using the error propagation technique. In figure 3, we show the constrained $w(z)$ at $68\%$ and $95\%$  confidence level. As one can see, the $w=-1$, cosmological constant behaviour is not always within the $95\%$  confidence region.  Within the redshift range $0.5\leq z \leq 1.25$, the $2\sigma$ confidence region does not allow the cosmological constant behaviour. This confirms the inconsistency of the cosmological constant model at $95\%$  confidence level. Moreover, the figure also confirms that models which are always non-phantom ($w>-1$) are ruled out at $95\%$  confidence level. This practically ruled out all canonical, minimally coupled scalar field models. Models which are always phantom ($w<-1$) or model where a phantom to non-phantom crossing happens are still consistent. This opens up the possibility for more exotic dark energy models.

\begin{figure*}[!htb]
\begin{center} 
\resizebox{200pt}{160pt}{\includegraphics{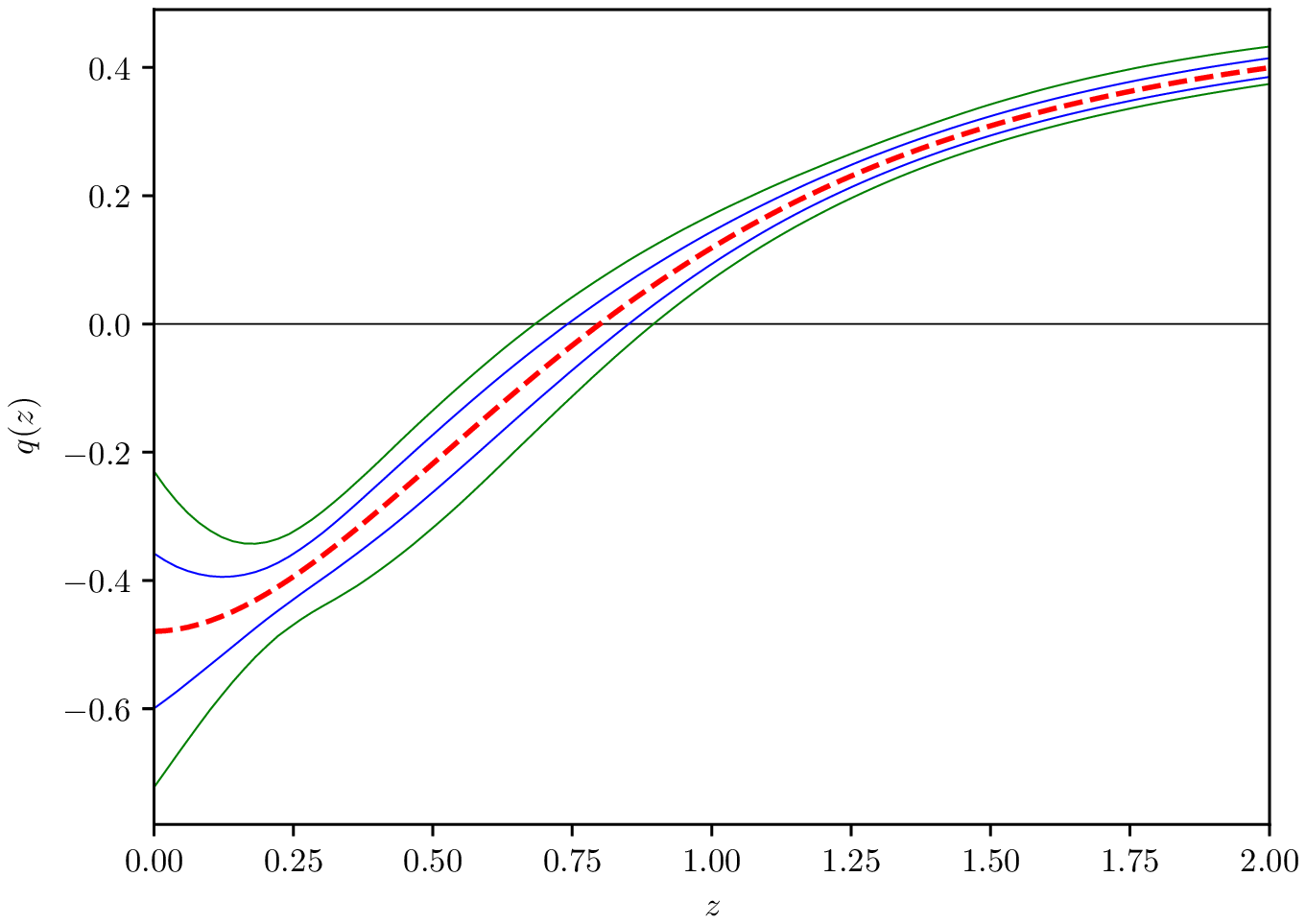}} 
\hspace{1mm} \resizebox{200pt}{160pt}{\includegraphics{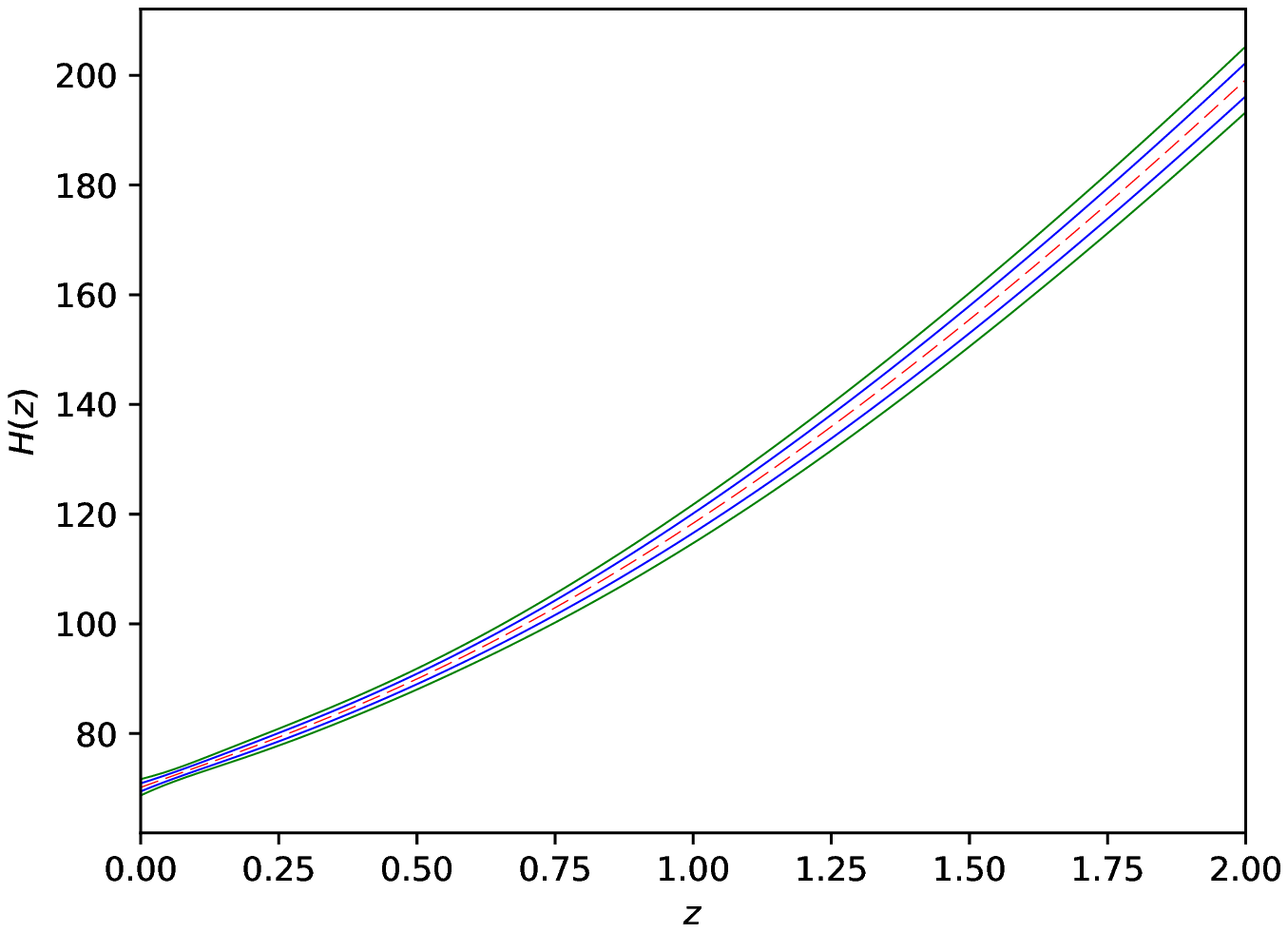}}
\end{center}
\caption{$68\%$ and $95\%$  confidence region for the reconstructed  deceleration parameter $q(z)$ and $H(z)$ for the same data set as in figure 2. The dashed line represent the best fit behaviour. } 
\end{figure*}
 
We also plot the reconstructed deceleration parameter $q(z)$ in figure 4. At $95\%$ confidence level, the universe starts accelerating approximately between $z=0.6$ and $z=0.8$. The reconstructed $H(z)$ is also shown in figure 4.

To check further whether the $\Lambda$CDM model is indeed inconsistent at $95\%$ confidence level, we further study the reconstructed behaviour of the $Om(z)$ parameter that we describe in equation  (4). As we mention above, for $\Lambda$CDM model the $Om$ parameter is constant throughout the evolution of the universe. In figure 5 (left panel), we show the reconstructed behaviour of $Om$ parameter as a function of redshift for CPL parametrisation. It is not difficult see that this reconstructed evolution of $Om$ parameter as a function of redshift does allow a constant behaviour at $95\%$confidence level making the $\Lambda$CDM model consistence at $95\%$ confidence level. 

To check whether our finding that a constant $Om$  is allowed at $95\%$ confidence level, depends on the dark energy parametrisation, we use a different parametrisation. This is recently proposed by Pantazis et al. \cite{7cpl} and is of the form ( we call it 7CPL in subsequent discussions):

\begin{equation}
w = w_{0} + w_{a} \left( \frac{z}{1+z}\right)^{7},
\end{equation}

\noindent
where $w_{0}$ and $w_{a}$ are the two parameters in this parametrisation. It is shown in \cite{7cpl} that while CPL parametrisation fits thawing model much better than the freezing model, the 7CPL parametrisation is an excellent fit to the freezing behaviour of dark energy. With such a different parametrisation, we again reconstruct the $Om$ parameter for the same set of observable data that we use for CPL. The reconstructed $Om(z)$ is shown in the right panel of figure 5. As one can see, although the overall behaviour for $Om(z)$ is different from that obtained for CPL parametrisation, a constant $Om$ is still allowed at $95\%$ confidence limit which confirms the consistency of the $\Lambda$CDM at $95\%$ confidence limit. This shows that the $\Lambda$CDM model is consistent with the current set observational data irrespective of the dark energy parametrisation.

We should mention that  unlike the equation of state of the dark energy, $Om$ does not depend explicitly on the current value of the $\Omega_{m0}$. The information on $H(z)$ is sufficient to reconstruct the $Om$. Although a  determination of $H$ from a single observable can suffer systematic uncertainties, but in our investigation, we use the determination of $H$ from SNIa, BAO, CMB acoustic scale, growth of matter fluctuations, as well independent measurement of $H$ from local experiments, to reconstruct the $Om(z)$. Hence this is a robust estimate for the $Om(z)$ parameter which clearly shows the consistency of the $\Lambda$CDM model with presently available observational data. One can also find out that for $\Lambda$CDM to be consistent,  the $95\%$ bound on $\Omega_{m0}$ for $\Lambda$CDM model is $[0.251, \hspace{1mm} 0.292]$ which is marginally lower than the Planck-2015 measurement of $\Omega_{m0}$ for $\Lambda$CDM model.  

\begin{figure*}[!htb]
\begin{center} 
\resizebox{200pt}{160pt}{\includegraphics{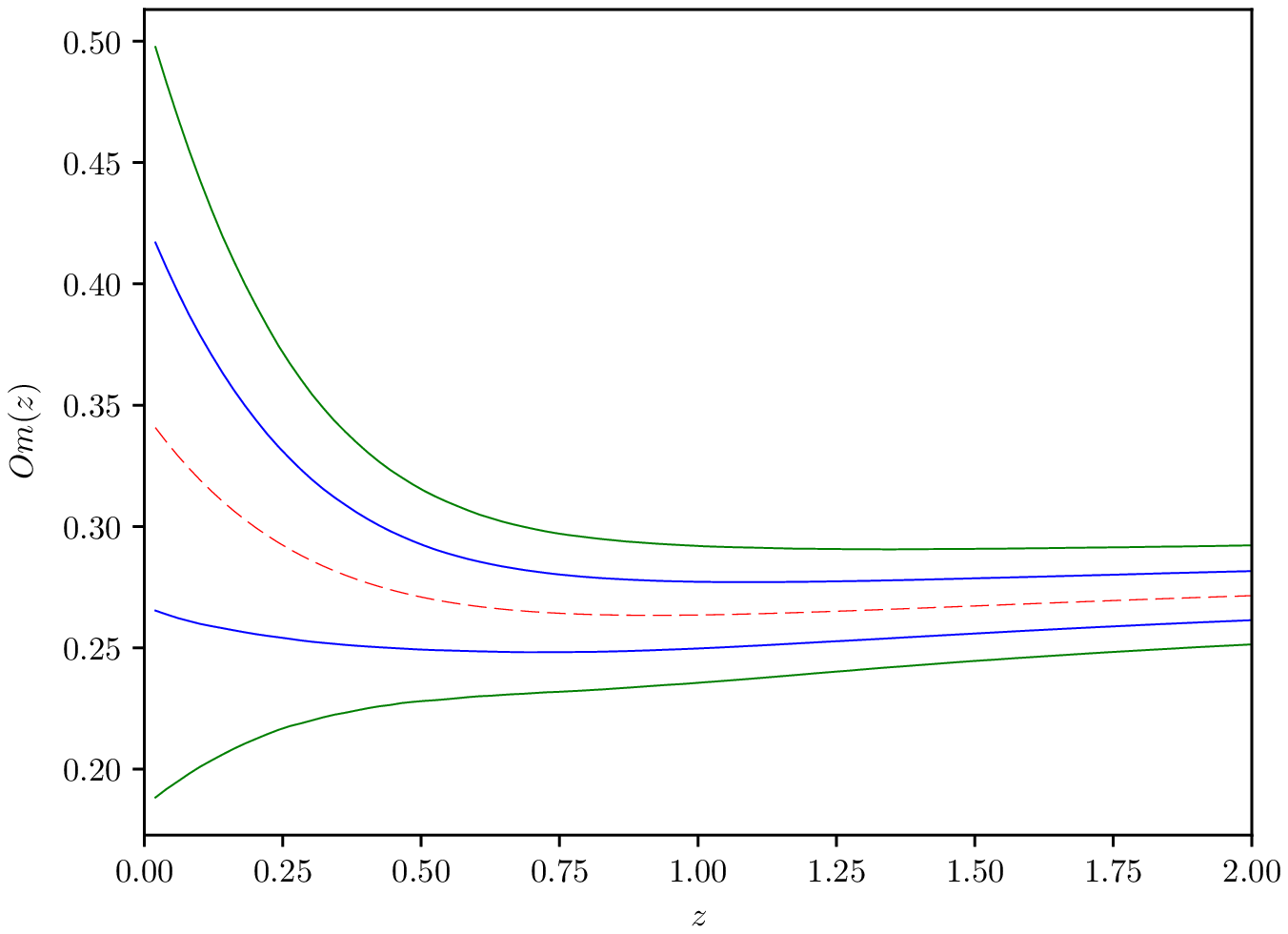}} 
\hspace{1mm} \resizebox{200pt}{160pt}{\includegraphics{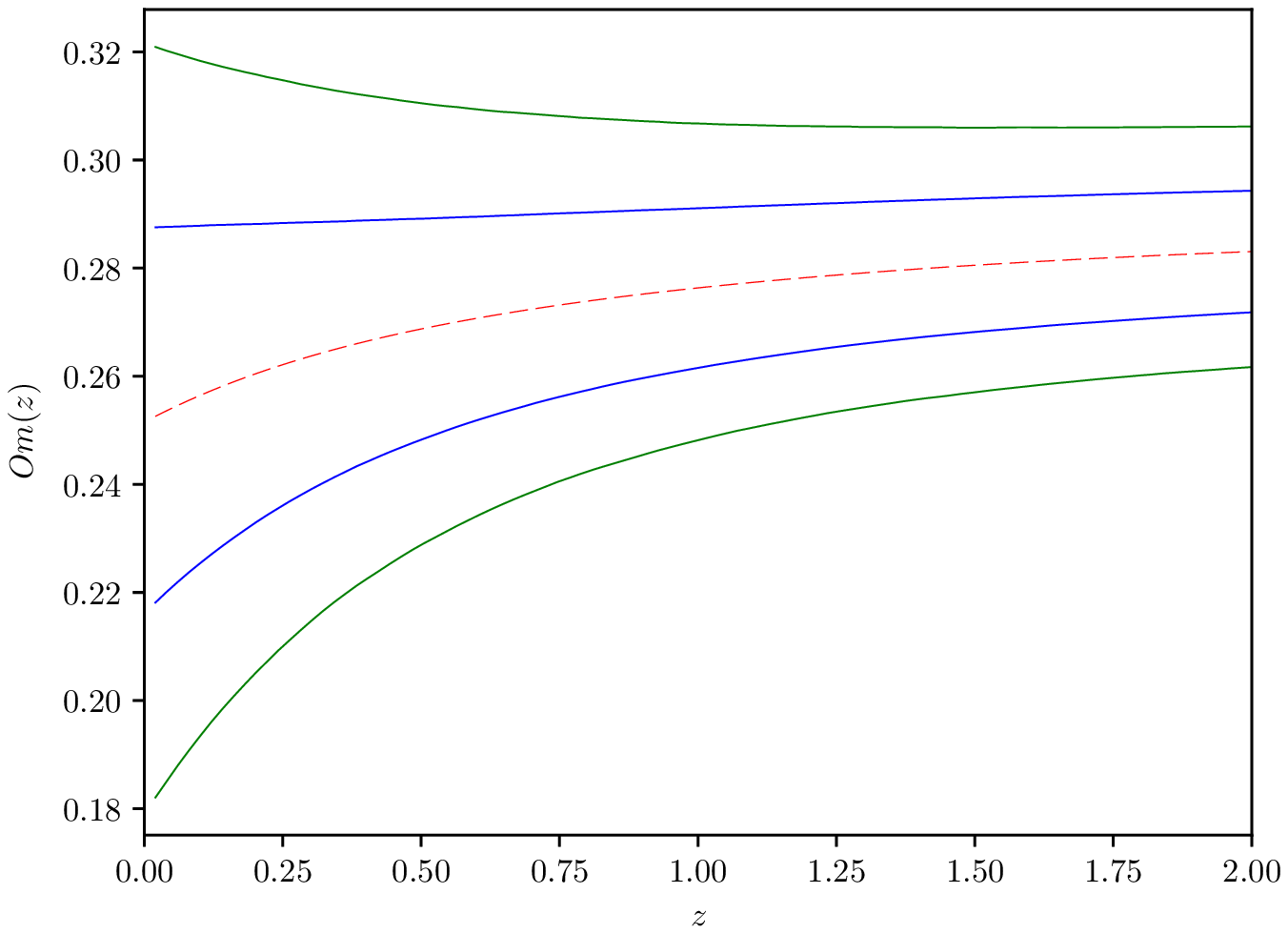}}
\end{center}
\caption{$68\%$ and $95\%$ confidence region for the reconstructed $Om(z)$. The dashed line represent the best fit behaviour. {\it Left}  figure is for CPL parametrisation and the {\it Right} figure is for 7CPL parametrisation. } 
\end{figure*}

To conclude, we have revisited the constraint on the evolution of the dark energy with a wide variety of presently available observational data. We use the CPL parametrisation to model the dark energy evolution and use the combination of SNIa, BAO, CMB shift and acoustic scale measurements, growth measurement as well as the measurement of $H((z)$ including the recent measurement of $H_{0}$. Recently with a combination of Planck2015+Riess16 dataset,Valentino et al \cite{valen} have shown that the $\Lambda$CDM model is inconsistent at $95\%$ confidence interval using the same parametrisation. We confirm this result using the reconstructed the equation of state of the dark energy with a different combination of datasets. Moreover the confidence contour in the $w_{0}-w_{a}$ parameters space also confirms the inconsistency of the freezing models at  $95\%$ confidence level whereas the thawing model is marginally allowed at $95\%$ confidence level.  But with our reconstructed $Om(z)$ diagnostic, we show that $\Lambda$CDM is indeed consistent with the current set of observational data with a slightly lower range of $\Omega_{m0}$ from that obtained by Planck-2015.We have also shown that measured value of the parameter $S = \sigma_{8} \sqrt{\Omega_{m}/0.3}$ as well as the confidence contour in the $\Omega_{m0}-\sigma_{8}$ plane is fully in agreement with the recent measurement by KiDS Survey. 

In most of the investigations to constrain the dark energy evolution, it is a standard practice to use the dark energy equation of state to obtain the constraint. But we should to be careful in this regard, because knowing the evolution of $H(z)$ from different data set, reconstruction of $w(z)$ for dark energy needs the information about $\Omega_{m0}$. The  $Om$ diagnostic is useful in this sense as the information about $\Omega_{m0}$ is not necessary for its reconstruction. Hence this diagnostic is a more powerful  {\it null test} for $\Lambda$CDM. Using this $Om$ diagnostic, we show that  the $\Lambda$CDM model is still very much consistent with the presently available data. We need to wait for some more time to see whether the dark energy is indeed $\Lambda$ or something else. Till that time, $\Lambda$ is still the best choice for the dark energy.

\vspace{5mm}
The author Ruchika thanks Council of Scientific and Industrial Research (CSIR), Govt. of India for Junior Research Fellowship. The author Anto I Lonappan thanks Center For Theoretical Physics, Jamia Millia Islamia, New Delhi, India for providing research facilities.

\end{document}